\documentclass[12pt]{article}
\usepackage{amsmath}
\newcommand{\III}{{\cal I}_3}
\newcommand{\IV}{{\cal I}_4}
\newcommand{\PP}{{\cal P}}

\begin{document}

\title{Pure quantum integrability}
\author{Jarmo Hietarinta\\
Department of Physics, University of Turku\\ FIN-20014 Turku, Finland}

\maketitle

\begin{abstract}
The correspondence between the integrability of classical mechanical
systems and their quantum counterparts is not a 1-1, although some
close correspondencies exist. If a classical mechanical system is
integrable with invariants that are polynomial in momenta one can
construct a corresponding commuting set of differential operators.
Here we discuss some 2- or 3-dimensional {\it purely} quantum
integrable systems (the 1-dimensional counterpart is the Lame
equation). That is, we have an integrable potential whose amplitude is
not free but rather proportional to $\hbar^2$, and in the classical
limit the potential vanishes. Furthermore it turns out that some of
these systems actually have $N+1$ commuting differential operators,
connected by a nontrivial algebraic relation. Some of them
have been discussed recently by A.P. Veselov {\it et. al.} from the
point of view of Baker-Akheizer functions.
\end{abstract}

\section{Introduction}
In classical mechanics the most common definition of integrability is
that of {\it Liouville integrability}: Suppose we have a system with $N$
degrees of freedom, that is, we have $N$ coordinates $q_i$ and $N$
conjugate momenta $p_j$, with Poisson brackets
$[q_i,p_j]=\delta_{ij}$.  This system is said to be {\em Liouville
integrable} \cite{Ar}, if there are $N$ functions $F_k(p,q)$ such that 1)
$[F_i,F_j]=0,\,\forall i,j$, and 2) the $F_i$ are functionally independent
and 3) sufficiently regular. One of the $F$'s is the Hamiltonian that
gives the dynamics. Weaker and stronger types of integrability have
also been used (e.g. partial integrability, algebraic integrability
and super-integrability).

When the idea of integrability is applied to quantum mechanics the
properties of commutativity and functional independence are mutually
exclusive: According to a theorem of von Neumann \cite{Neu}, if a pair
of self-adjoint operators $A$ and $B$ commutes, then there is a third
self-adjoint operator $C$ such that $A=f(C)$, $B=g(C)$. This problem
is avoided if we concentrate on differential operators, which anyway
are the most interesting from practical point of view. Thus, even if
$A=i\partial_x$ and $B=i\partial_y$ commute, the operator $C$
mentioned above is not a differential operator.  In this paper we
consider only differential operators and property 3) above is then
automatically satisfied, but we should relax the independence
requirement 2): we should only ignore cases where one of the commuting
operator can be given as a polynomial of the others.

In this article we study the existence of operators commuting with a
standard type two-dimensional Hamiltonian (Schr\"odinger operator)
\begin{equation}
{\cal H}=-\tfrac12{\hbar^2}\left(\partial_x^2+\partial_y^2\right)+V(x,y).
\label{Ham2d}
\end{equation}
This may be considered as arising from the classical system
$H=\frac12(p_x^2+p_y^2)+V(x,y)$ by the usual correspondence $p_\alpha
\to -i\hbar\partial_\alpha$. (The classical integrability of such
systems has been reviewed in \cite{JH87}.)  The transition from
classical to quantum integrability seems to be always possible,
sometimes one just has to add correction terms to the Hamiltonian and
to the operator commuting with it, but these correction terms are
always $O(\hbar^2)$ \cite{JHq}. Recently the existence of a
differential operator commuting with (\ref{Ham2d}) has been discussed
at length in \cite{OO} (along with multi-dimensional
generalizations). In that work strong assumptions were made on the
symmetries of the potential (invariance under the Weyl group), but it
turns out that the purely quantum integrable (PQI) systems discussed
here do not have such symmetry properties.

An important difference between classical and quantum integrability
lies in the question of independence as mentioned above. Thus it will
be meaningful and interesting to have $N+1$ commuting operators, even
if they must be algebraically related (sometimes this is called {\it
algebraic integrability}). This can be already illustrated for
1-dimensional systems. They are always integrable (the Hamiltonian
commutes with itself) but some systems are more integrable that
others. For example \cite{JHs} the Lame operator
\begin{equation}
{\cal L}=-\tfrac12\hbar^2\dfrac{d^2}{dx^2}+\hbar^2\PP(x),
\end{equation}
where $\PP$ is the Weierstrass elliptic function, commutes with
\begin{equation}
{\cal I}=-i\hbar^3\frac{d^3}{dx^3}+\tfrac32i\hbar^3
\bigl\{\frac{d}{dx},\PP(x)\bigr\},
\end{equation}
where the bracket $\{.,.\}$ stands for an anticommutator.
The operators ${\cal L}$ and ${\cal I}$ are algebraically related,
\begin{equation}
{\cal I}^2=8{\cal L}^3-\tfrac12\hbar^4g_2{\cal L}+\tfrac14\hbar^6g_3,
\end{equation}
(where $g_i$ are the constants characterizing $\cal P$) and this
relation can be used to express the wave-function in terms of other
elliptic functions. The potential $\hbar^2\PP(x)$ is the only one that
has a third order commuting operator, for a fifth order operator
generalizations are possible \cite{JHs}.

The classification of commuting ordinary differential operators was
studied already in the 1920's \cite{BC} and more recently this problem
has been studied from the point of view of soliton theory. The
corresponding two dimensional problem (i.e., the existence of three
commuting differential operators) was studied in \cite{CV}. In the
one-dimensional case the algebraically integrable cases are (trivially)
special cases of normal (classical) integrability, and in \cite{CV} it
was conjectured that this holds also in higher dimensions, but it
turns out not to be the case.

In Section 2 we present results from a search for a third order
differential operator, commuting with $\cal H$ of (\ref{Ham2d}). We
find that there are indeed some PQI systems, whose potential is
proportional to $\hbar^2$ and therefore vanishes in the classical
limit.  In section 3 we show, that the integrable systems found in
Section 2 also have a fourth order commuting operator, and that the
three operators are polynomially related. In Section 4 we consider
fourth order differential operators commuting with the Hamiltonian
(\ref{Ham2d}) with a potential that is a generalization of the
rational of four term Calogero-Moser potential. Again some PQI
potentials are found.

\section{Third order commuting operator}
We will consider only those third order differential operators $\III$
whose leading order part has constant coefficients. The analysis of
the leading terms is the same as that of the corresponding classical
system; this follows from the Weyl correspondence and the fact that
the leading part of the Moyal bracket (which represents the
commutator) agrees with the Poisson bracket \cite{JHq}.  Since the
second order operator ${\cal H}$ has no first order derivatives, we
may assume that $\III$ has no second order derivatives, i.e.,
\begin{equation}
\III=(i\hbar)^3(c_3 \partial_x^3+c_2\partial_x^2\partial_y^{\phantom
2} +c_1\partial_x^{\phantom 2}\partial_y^2+c_0 \partial_y^3)
+d_{11}\partial_x+ d_{10}\partial_y+d_0.
\end{equation}
From the leading terms of the commutation condition $[\III,{\cal
H}]=0$ one immediately derives a linear PDE for the potential $V$
\begin{equation}
c_1V_{xxx}+(3c_0-2c_2)V_{xxy}+(3c_3-2c_1)V_{xyy}+c_2V_{yyy}=0,
\end{equation}
where the subscripts stand for partial derivatives. The generic
solution of this equation is
\begin{equation}
V=g_1(\alpha_1x+\beta_1y)+g_2(\alpha_2x+\beta_2y)+g_3(\alpha_3x+\beta_3y),
\label{V1}
\end{equation}
where $(\alpha_i,\beta_i)$ are the three solutions of $c_1\alpha^3+
(3c_0-2c_2)\alpha^2\beta+ (3c_3-2c_1)\alpha\beta^2 +c_2\beta^3=0$.
(The non-generic cases need separate study.)

The potential (\ref{V1}) is form invariant under coordinate rotations,
which just imply changes in the parameters $\alpha_i,\beta_i$. This
suggests writing all formulae in a rotationally covariant form.
Furthermore, we may assume that the parameters $\alpha_i,\beta_i$
satisfy the condition
\begin{equation}
\alpha_1+\alpha_2+\alpha_3=0,\quad\beta_1+\beta_2+\beta_3=0.
\label{parlin}
\end{equation}
(To obtain this it is only necessary to scale the arguments of the
functions $g$.) The sub-manifold (\ref{parlin}) is invariant under
parameter rotations. Instead of $x,y$ and the corresponding partial
derivatives let us use the quantities
\begin{equation}
X_i=\alpha_i\,x+\beta_i\,y,\, \check X_i=\beta_i\, x- \alpha_i\, y,
\quad L_i= -i\hbar(\beta_i\,\partial_x - \alpha_i\,\partial_y),
\, \check L_i= -i\hbar(\alpha_i\,\partial_x + \beta_i\,\partial_y),
\end{equation}
which satisfy
\begin{equation}
[L_i,X_j]=-[\check L_i,\check X_j]=i\hbar D_{ij},\quad
[L_i,\check X_j]=[\check L_i,X_j]=-i\hbar\Delta_{ij},
\end{equation}
where
\begin{equation}
D_{ij}:=\alpha_i\beta_j-\alpha_j\beta_i,\quad
\Delta_{ij}:=\alpha_i\alpha_j+\beta_i\beta_j.
\end{equation}
Due to (\ref{parlin}) we have $D\equiv D_{12}= D_{23}= D_{31}=
 -D_{21}= -D_{32}= -D_{13}$.
On the manifold (\ref{parlin}) we have $\Delta_{1i}+ \Delta_{2i}+
\Delta_{3i} =0$ so that we have the following relations between the
diagonal and non-diagonal elements:
\begin{equation}
\Delta_{ii}=-(\Delta_{ij}+\Delta_{ki}),\quad
\Delta_{ij}=\tfrac12(-\Delta_{ii}-\Delta_{jj}+\Delta_{kk}),
\label{ndiag}
\end{equation}
where $\{i,j,k\}$ is a permutation of $\{1,2,3\}$. Furthermore
\begin{equation}
L_1+L_2+L_3=0, \quad
L_1^2\Delta_{23}+L_2^2\Delta_{31}+L_3^2\Delta_{12}=
\hbar^2 D^2(\partial_x^2+\partial_y^2).
\end{equation}

By a direct computation we find the following expression for $\III$
\begin{eqnarray}
\III&=&L_1^3\Delta_{23}^2+L_2^3\Delta_{31}^2+L_3^3\Delta_{12}^2
\nonumber\\
&&+\tfrac32D^2\sum_{j=1}^3\left\{
L_j,(\sum_{k=1}^3 g_k\Delta_{jk})-2g_j\Delta_{jj}\right\},
\end{eqnarray}
but there is one remaining condition that can be written as
\begin{eqnarray}
&&(\Delta_{22}-\Delta_{33}) (g_1''' \hbar^2 \Delta_{11} - 12 g_1'g_1)
\nonumber\\
&+&(\Delta_{33}-\Delta_{11}) (g_2''' \hbar^2 \Delta_{22} - 12 g_2'g_2)
\label{s3eq}\\
&+&(\Delta_{11}-\Delta_{22}) (g_3''' \hbar^2 \Delta_{33} - 12 g_3' g_3)
\nonumber\\
&-&12\left[
(g_1g_2'-g_1'g_2)\Delta_{12}+
(g_2g_3'-g_2'g_3)\Delta_{23}+
(g_3g_1'-g_3'g_1)\Delta_{31}\right]=0.\nonumber
\end{eqnarray}

The complete solution of (\ref{s3eq}) is not known, but several
solutions can be found when we note that the Weierstrass elliptic
function $\PP$ satisfies the equations
\begin{equation}
\PP'''=12 \PP \PP',\qquad \begin{vmatrix}1 & \PP(u) & \PP(u)'\\ 1 &
\PP(v) & \PP(v)'\\1 & \PP(w) & \PP(w)'\end{vmatrix}=0,\mbox{ where
}u+v+w=0.
\end{equation}
Thus if we assume that $g_i=a_i\PP$ we can identify the last three
terms of (\ref{s3eq}) as the above determinant, if
\begin{equation}
a_1 a_2\Delta_{12}= a_2 a_3\Delta_{23}= a_3
a_1\Delta_{31}.\label{eqs32}
\end{equation}
The first three terms of (\ref{s3eq}) imply three further equation:
\begin{eqnarray}
(\Delta_{22}-\Delta_{33})
(\hbar^2 \Delta_{11} - a_1)&=&0,\nonumber\\
(\Delta_{33}-\Delta_{11})
(\hbar^2 \Delta_{22} - a_2)&=&0,\label{eqs33}\\
(\Delta_{11}-\Delta_{22})
(\hbar^2 \Delta_{33}-a_3)&=&0.\nonumber
\end{eqnarray}

From (\ref{eqs32}) we immediately find that
\begin{equation}
a_i=\Delta_{jk}X,\, \mbox{ ($i,j,k$ cyclic)},
\label{amp}
\end{equation}
so that in any case the potential will be of the form
\begin{equation}
V\propto \Delta_{23}\PP(\alpha_1x+\beta_1y)+
\Delta_{31}\PP(\alpha_2x+\beta_2y)+\Delta_{12}\PP(\alpha_3x+\beta_3y).
\label{genppot}
\end{equation}
The solutions of (\ref{amp},\ref{eqs33}) can be divided into three
groups:

\subsubsection*{Case 1: All $\Delta_{ii}$'s equal.}
From (\ref{ndiag}) we find that the off-diagonal $\Delta$'s and
therefore $a_i$'s are all equal. The equality of $\Delta_{ii}$'s
allows the parameterization $\alpha_i=2A\cos(\theta_i),\,
\beta_i=2A\sin(\theta_i)$, and from (\ref{parlin}) it follows that
$\theta_2= \theta_1+ \tfrac23\pi,\, \theta_3= \theta_1-\tfrac23\pi$.
This is the well known solution with a nonzero classical limit. In
addition to the overall rotation there is the freedom of the
amplitudes $A$ and $a$. If we choose $\theta_1=\pi/6$ we get the
familiar form
\begin{equation}
V_{3.1}=a\bigl[\PP(A({\sqrt 3}x+y))+ \PP(A(-{\sqrt 3}x+y))+\PP(-2Ay)\bigr].
\end{equation}

\subsubsection*{Case 2: Two $\Delta_{ii}$'s equal.}
Let us say $\Delta_{11}=\Delta_{22}\neq \Delta_{33}$.  Then from
(\ref{eqs33}) $X= \hbar^2\Delta_{11}/\Delta_{23}=
-2\hbar^2\Delta_{11}/\Delta_{33}$ and $a_1=a_2=\hbar^2 \Delta_{11},
a_3=\hbar^2 \Delta_{11}(2\Delta_{11}- \Delta_{33})/ \Delta_{33}$. Note
the overall $\hbar^2$ in the amplitudes $a_i$. One possible
parameterization (after fixing the rotational freedom) is
\begin{equation}
V_{3.2}=\hbar^2\frac{A^2+B^2}{2A^2}\bigl[2A^2\PP(Bx+Ay)+2A^2\PP(-Bx+Ay)+
(B^2-A^2)\PP(-2Ay)\bigr].
\label{V32}
\end{equation}
This is the solution found by Veselov {\it et. al.} \cite{VFC,APV} for the
algebraic special case $\PP(x)=x^{-2}$.

\subsubsection*{Case 3: All $\Delta_{ii}$'s different.}
Then we must have $a_i=\hbar^2 \Delta_{ii}$ by (\ref{eqs33}), and
(\ref{amp}) implies $\Delta_{11}+\Delta_{22}+ \Delta_{33} = 0$. Thus
some of the parameters must be complex.  One parameterizations is
given by
\begin{align}
V_{3.3}=\hbar^2A^2\bigl[&
2(-1+i\sqrt3\sin(\theta))
\PP(i\sqrt3A\cos(\theta)x+A(1+i\sqrt3\sin(\theta))y)\nonumber\\
&+2(-1-i\sqrt3\sin(\theta))
\PP(-i\sqrt3A\cos(\theta)x+A(1-i\sqrt3\sin(\theta))y)\nonumber\\
&+4\PP(-2Ay)\bigr].
\label{V33}
\end{align}
Note that even though some $\alpha_i,\beta_i$ must be complex, it is
possible to get a real potential $V$, e.g., if $\PP$ itself is real.
This solution seems to be new.

Note that Case 2 intersects with Case 1 at $B=\sqrt{3}A$
($a=4\hbar^2A^2$), and with Case 3 at $B=i\sqrt{3}A$
($\theta=0$ in (\ref{V33})). At these special points the system
should have more symmetries.

For Cases 2 and 3 the amplitudes are proportional to $\hbar^2$
so they are PQI. Although the amplitudes are fixed, there are two
nontrivial degrees of freedom as in case 1, but they are now all in
the direction parameters $\alpha_i,\beta_i$.

\subsection{Connection with a three dimensional system}
A Hamiltonian of the form (\ref{Ham2d},\ref{V1}) (with the term
$-\Delta_{12}\Delta_{23}\Delta_{31}\,p_z^2/D^2$ added) can also be
written as a three dimensional system
\begin{equation}
\hskip -0.5cm H=-\frac{\hbar^2}2\left(\Delta_{23}\,\partial_1^2
+\Delta_{31}\,\partial_2^2+ \Delta_{12}\,\partial_3^2\right)
+g_1(q_2-q_3)+g_2(q_3-q_1)+g_3(q_1-q_2),
\end{equation}
through the linear transformation
\begin{equation}\left\{\begin{array}{rcl}
x&=&-(q_1 \beta_1+q_2 \beta_2+q_3 \beta_3)/D,\\
y&=& (q_1 \alpha_1+q_2 \alpha_2+q_3 \alpha_3)/D,\\
z&=& (q_1 \Delta_{13}\Delta_{12}+ q_2 \Delta_{23}
\Delta_{12}+ q_3 \Delta_{13} \Delta_{23})/D^2,\end{array}\right.
\end{equation}
and correspondingly
\begin{equation}\left\{\begin{array}{rcl}
\partial_x&=&(\partial_1 \beta_1 \Delta_{23}+\partial_2
\beta_2 \Delta_{13} +\partial_3 \beta_3 \Delta_{12}) /D,\\
\partial_y&=&-(\partial_1 \alpha_1 \Delta_{23}+\partial_2
\alpha_2\Delta_{13} +\partial_3 \alpha_3 \Delta_{12})/D,\\
\partial_z&=&\partial_1+\partial_2+\partial_3.
\end{array}\right.
\end{equation}

If we denote $\mu_i=\Delta_{jk}$, ($i,j,k$ cyclic) and use
(\ref{genppot}) the results can combined as
\begin{equation}
H=-\tfrac{\hbar^2}{2}\left(\mu_1\partial_1^2+\mu_2\partial_2^2+
\mu_3\partial_3^2\right) +A\bigl[\mu_1\PP(q_2-q_3)+\mu_2\PP(q_3-q_1)+
\mu_3\PP(q_1-q_2)\bigr],
\end{equation}
and the three integrable cases are characterized as follows:
\begin{enumerate}
\item The classical case: $\mu_1=\mu_2=\mu_3$, $A$ free.
\item Elliptic PQI case: $\mu_1=\mu_2$,
$A=-\hbar^2(\mu_1+\mu_3)/\mu_1$
\item Hyperbolic PQI case: $\mu_1+\mu_2+\mu_3=0$,
$A=\hbar^2$.
\end{enumerate}
Note that for the two PQI cases the same expression $A=\hbar^2[1-
(\mu_1+\mu_2+\mu_3)/\mu_1]$ works for the overall amplitude.

In \cite{VFC,APV} the $n$-dimensional elliptic case was associated to
a non-Coxeter configuration called ${\cal A}_n(m)$, similar geometric
interpretation for the hyperbolic case would be interesting.

\section{Fourth order commuting operator}
In this section we consider the operator
\begin{equation}
\IV=(i\hbar)^4\,\sum_{i=0}^4c_i\partial_x^i\partial_y^{4-i}+\dots
\end{equation}
and now the condition of commutativity $[\IV,{\cal H}]=0$ leads to the
necessary condition
\begin{equation}
c_1V_{xxxx}+2(2c_0-c_2)V_{xxxy}+3(c_3-c_1)V_{xxyy}
+2(c_2-2c_4)V_{xyyy}-c_3V_{yyyy}=0.
\label{V4}
\end{equation}
Although this is a fourth order equation it does not allow four
arbitrary directions $(\alpha_i,\beta_i)$, because of the way the
$c_i$'s enter in the equation. We will return to this problem later
and first consider potential with three terms.

\subsection{Three term potential}
By direct calculation with (\ref{V1},\ref{parlin}) one finds that the
invariant can be written as
\begin{eqnarray}
\IV&=&L_1^4\Delta_{23}^3+L_2^4\Delta_{31}^3+L_3^4\Delta_{12}^3\nonumber\\
&&-D^2\sum_{j,k=1}^3 \left\{L_j^2,-g_j\Delta_{j+1,j+2}^2
+g_k(\Delta_{k,j+1}\Delta_{k,j+2}+3\Delta_{j+1,j+2}^2)\right\}
+C(x,y).
\end{eqnarray}
Here the indices for $\Delta$ are to be taken modulo 3. The
integrability condition for $C$ can be written as
\begin{equation}
(\Delta_{23}\check L_1+\Delta_{31}\check L_2+\Delta_{12}\check L_3)\,\Omega=0
\end{equation}
where
\begin{eqnarray}
\Omega&=&K_1(\Delta_{22}-\Delta_{33}) (g_1''' \hbar^2 \Delta_{11} - 12
g_1'g_1) \nonumber\\ &+&K_2(\Delta_{33}-\Delta_{11}) (g_2''' \hbar^2
\Delta_{22} - 12 g_2'g_2)
\label{s4eq}\\
&+&K_3(\Delta_{11}-\Delta_{22}) (g_3''' \hbar^2 \Delta_{33} - 12 g_3'
g_3) \nonumber\\ &-&12\left[ (g_1g_2'-g_1'g_2)\Delta_{12}+
(g_2g_3'-g_2'g_3)\Delta_{23}+
(g_3g_1'-g_3'g_1)\Delta_{31}\right],\nonumber
\end{eqnarray}
and
\begin{equation*}
K_i=\frac{3\Delta_{i,i+1}\Delta_{i,i+2}}
{\Delta_{ii}\Delta_{i+1,i+2}+2\Delta_{i,i+1}\Delta_{i,i+2}}
\end{equation*}
(The denominator of $K_i$ can also be written as $[\Delta_{23}\check
L_1+\Delta_{31}\check L_2+\Delta_{12}\check L_3,\check X_i]$.)  The
expression (\ref{s4eq}) is the same as (\ref{s3eq}), except for the
extra coefficients $K_i$, which however play no role if (\ref{s4eq})
is solved by $g_i=a_i\PP$ with (\ref{eqs32},\ref{eqs33}).  The
integrability condition for $C$ may have other solutions, but we can
say at least that all solutions obtained in Section 2 also have a
fourth order invariant. However, since ${\cal H}^2$ is also a fourth
order invariant commuting with $\cal H$ we must discuss the
independence of the newly found invariants.

\subsection{Relations between the invariants}
Since the system is two dimensional there can be at most two
algebraically independent commuting quantities. Indeed on finds
relations between the three operators found above, for example in Case
1 we have
\begin{equation}
\IV=-\Delta_{11} D^4 {\cal H}^2,
\label{E:v31rel}
\end{equation}
so that in this case the new fourth order invariant is useless.

For the other cases the algebraic relation is less trivial and
therefore the extra invariant provides additional information. The
computations are rather extensive and we have verified only the
relation connecting the leading terms of ${\cal I}_j$, denoted below
by $I_j$. The results are as follows:

For case 2 when $\Delta_{22}=\Delta_{11}$ we have the relation
\begin{equation}
\begin{split}
&I_2^6 \Delta_{33}^3 (16 \Delta_{11}^4 + 16 \Delta_{11}^3 \Delta_{33}
- 20 \Delta_{11}^2 \Delta_{33}^2 - 12 \Delta_{11} \Delta_{33}^3 + 9
\Delta_{33}^4)\\& + 12 I_2^4 I_4 \Delta_{33}^2 (16 \Delta_{11}^4 + 8
\Delta_{11}^3 \Delta_{33} - 16 \Delta_{11} ^2 \Delta_{33}^2 - 2
\Delta_{11} \Delta_{33}^3 + 3 \Delta_{33}^4)\\&  + 32 I_2^3 I_3^2
\Delta_{33}^2 (12 \Delta_{11}^4 - 26 \Delta_{11}^3 \Delta_{33} + 15
\Delta_{11}^2 \Delta_{33}^2 - \Delta_{33}^4)\\&  + 48 I_2^2 I_4^2
\Delta_{11} \Delta_{33} (16 \Delta_{11} ^3 - 9 \Delta_{11}
\Delta_{33}^2 + 2 \Delta_{33}^3)\\&  + 384 I_2 I_3^2 I_4 \Delta_{11}
\Delta_{33} (4 \Delta_{11}^3 - 9 \Delta_{11} ^2 \Delta_{33} + 6
\Delta_{11} \Delta_{33}^2 - \Delta_{33}^3)\\&  + 256 I_3^4 \Delta_{11} (4
\Delta_{11}^4 - 13 \Delta_{11}^3 \Delta_{33} + 15 \Delta_{11}^2
\Delta_{33}^2 - 7 \Delta_{11} \Delta_{33}^3 + \Delta_{33}^4)\\&  + 64
I_4^3 \Delta_{11}^2 (16 \Delta_{11}^2 - 8 \Delta_{11} \Delta_{33} +
\Delta_{33}^2)=0,
\end{split}
\label{rel24}
\end{equation}
which is of degree 12 in the derivatives. If $\Delta_{11}=\Delta_{33}$
(intersection with Case 1) this expression simplifies to $9\Delta_{11}^4
(\Delta_{11}I_2^2+4I_4)^3=0$, in agreement with (\ref{E:v31rel}).
It simplifies also if $\Delta_{33}=4 \Delta_{11}$ but this corresponds
to the trivial limit $B=0$.

In case 3 we get the following complicated identity among the leading
terms:
{\allowdisplaybreaks
\begin{align}
&\,i_4^6 (a^2-a+1)^4 a^2 (a-1)^2\nonumber\\
-&6 \,i_4^4 \,i_2^2  (a^2-a+1)^3 a (a-1) (a^6-3a^5+5a^3-3a+1)\nonumber\\
+& 12 \,i_4^4 \,i_3^2 \,i_2
(a^2-a+1)^3 a (a-1) (2a^6-6a^5+3a^4+4a^3+3a^2-6a+2)\nonumber\\
+&3 \,i_4^4 \,i_2^4  (a^2-a+1)^2
  (3 a^{12} - 18 a^{11} + 19 a^{10} + 70 a^9 - 140 a^8 - 58 a^7 +
    251 a^6\nonumber\\
&\hskip2cm - 58  a ^5 - 140 a^4 + 70 a^3 + 19 a^2 - 18 a + 3)\nonumber\\
-&2 \,i_4^3 \,i_3^4 (a^2-a+1)^3 a (a-1)
 (8 a^6 - 24 a^5 + 21 a^4 - 2 a^3 + 21 a^2 - 24 a + 8)\nonumber\\
-&4 \,i_4^3 \,i_3^2 \,i_2^3 (a^2-a+1)^2
 (2 a^{12} - 12 a^{11} - 44 a^{10} + 330 a^9 - 425 a^8 - 412 a^7 +
    1124 a^6\nonumber\\
&\hskip2cm - 412 a^5 - 425 a^4 + 330 a^3 - 44 a^2 - 12 a + 2)\nonumber\\
+&2 \,i_4^3 \,i_2^6 (a^2-a+1) a (a-1)
 (35 a^{12} - 210 a^{11} + 283 a^{10} + 510 a^9 - 1268 a^8 - 298 a^7
 \nonumber\\&\hskip2cm +
   1931 a^6 - 298 a^5 - 1268 a^4 + 510 a^3 + 283 a^2 - 210 a + 35)\nonumber\\
-&18 \,i_4^2 \,i_3^4 \,i_2^2  (a^2-a+1)^2 a^2 (a-1)^2
(a^2 - a - 5)(5 a^2 - 11 a + 5)(5 a^2 + a - 1)\nonumber\\
-&12 \,i_4^2 \,i_3^2 \,i_2^6 (a^2-a+1) a (a-1)
  (13 a^{12} - 78 a^{11} + 54 a^{10} + 445 a^9
 - 786 a^8 - 384 a^7\nonumber\\&\hskip2cm +
   1485 a^6 - 384 a^5 - 786 a^4 + 445 a^3 + 54 a^2 - 78 a + 13)\nonumber\\
+&3 \,i_4^2 \,i_2^8  (2 a^{18} - 18 a^{17} + 112 a^{16}
 - 488 a^{15} + 1012 a^{14} + 28 a^{13} -
 3599 a^{12} + 4330 a^{11}\nonumber\\&\hskip2cm
 + 2433 a^{10} - 7622 a^9 + 2433 a^8 +
 4330 a^7 - 3599 a^6 + 28  a^5 + 1012 a^4\nonumber\\
&\hskip2cm  - 488 a^3 + 112 a^2 -
 18 a + 2)\nonumber\\
+&108 \,i_4 \,i_3^6 \,i_2 (a^2-a+1)^2 a^2 (a-1)^2
(a^2 - 4 a + 1)(a^2 + 2 a - 2)(2 a^2 - 2 a - 1)\nonumber\\
+&6 \,i_4 \,i_3^4 \,i_2^4  (a^2-a+1) a (a-1)
(8 a^{12} - 48 a^{11} - 120 a^{10} + 1040 a^9
 - 1323 a^8 - 1476 a^7\nonumber\\&\hskip2cm +
 3846 a^6 - 1476 a^5 - 1323 a^4 + 1040 a^3 - 120 a^2 - 48 a + 8)\nonumber\\
-&12 \,i_4 \,i_3^2 \,i_2^7
(a^3 - 3 a + 1)(a^3 - 3 a^2 + 1) (2 a^{12} - 12 a^{11} + 41 a^{10} -
95 a^9 + 86 a^8 + 94 a^7\nonumber\\
&\hskip2cm - 230 a^6 + 94  a^ 5 + 86 a^4 - 95 a^3
+ 41 a^2 - 12 a + 2)\nonumber\\
+&6 \,i_4 \,i_2^{10}
(a - 2)^2( a + 1)^2( 2 a - 1)^2( a^2 - a + 1)^2 ( a - 1)( a^3 -
2 a^2 - a + 1)( a^3 - a^2 - 2 a + 1)\nonumber\\
+&729 \,i_3^8 a^4(a-1)^2 (a^2-a+1)^2\nonumber\\
+&4 \,i_3^6 \,i_2^3 (a^2-a+1) a (a-1)
(8 a^{12} - 48 a^{11} + 195 a^{10} - 535 a^9
+ 342 a^8 + 1314 a^7\nonumber\\&\hskip2cm  -
2544 a^6 + 1314 a^5 + 342 a^4 - 535 a^3 + 195 a^2
 - 48 a + 8)\nonumber\\
+&2 \,i_3^4 \,i_2^6
(8 a^{18} - 72 a^{17} + 231 a^{16} - 216 a^{15}
 - 777 a^{14} + 3507 a^{13} -
6201 a^{12} + 3003 a^{11}\nonumber\\
&\hskip2cm  + 7470 a^{10} - 13898 a^9 + 7470 a^8 +
3003 a^7 - 6201 a^6 + 3507 a^5 - 777 a^4\nonumber\\
&\hskip2cm  - 216 a^3 + 231 a^2 -
72 a + 8)\nonumber\\
-&4 \,i_3 \,i_2^9
(a^2 - a + 1)^2
a(a - 1)( a^3 - 3 a + 1)( a^3 - 3 a^2 + 1)\nonumber\\
&\hskip2cm  ( 5 a^6 - 15 a^5 -
3 a^4 + 31 a^3 - 3 a^2 - 15 a + 5)\nonumber\\
+& \,i_2^{12} (a^2-a+1)^4(a^3 - 2 a^2 - a + 1)^2
(a^3 - a^2 - 2 a + 1)^2=0,\label{rel34}
\end{align}}
where we have used the parameterization
$\Delta_{11}=(1-a)b,\,\Delta_{22}=ab,\, \Delta_{33}=-b,$ and $i_j=
b^{1-j} I_j$.

The conditions of Cases 2 and 3 intersect when
$\Delta_{ii}=\Delta_{jj}=-2\Delta_{kk}$. The various permutations of
this correspond to $a=1/2, -1$ and $a=2$ and  for these special values
(\ref{rel34}) becomes a
square of (\ref{rel24}), for $a=-1$ and $2$
we get the relation
\begin{equation}
\,i_2^6 + 12 \,i_2^3 \,i_3^2 - 3 \,i_2^2 \,i_4^2 + 36 \,i_2 \,i_3^2
\,i_4 - 36 \,i_3^4 + 2 \,i_4^3=0
\end{equation}
and for $a=1/2$ (i.e., $\mu_1=\mu_2=-\tfrac12\mu_3$)
\begin{equation}
\,i_2^6 - 24 \,i_2^3 \,i_3^2 - 12 \,i_2^2 \,i_4^2 + 144 \,i_2 \,i_3^2
\,i_4 - 144 \,i_3^4 - 16 \,i_4^3=0
\end{equation}
(The second form is obtained from the first by $i_3\to \sqrt{-2} i_3$,
$i_4\to -2 i_4$.)  In fact one can show that the expression
(\ref{rel34}) can be written as $U^2+(a+1)^2(a-1/2)^2(a-2)^2 V=0$,
where $U$ and $V$ are some polynomials of $a$ and the $i_j$'s.
Whether (\ref{rel34}) factorizes also for some other special values is
an open question.

\subsection{Four term potential}
Let us now return to equation (\ref{V4}). Since it is a fourth order
linear equation is should have solutions of type $V=\sum_{i=1}^4
g(\alpha_i x+\beta_i y)$. This is indeed the case, but since we really
have only four rather than five $c_i$'s at our disposal (we can use
${\cal H}^2$ to eliminate $c_2$, say) there will be a relation between
$\alpha_i,\,\beta_i$. In order to simplify subsequent computations we
will use rotational invariance to fix $\alpha_1=0,$ and scale so that
$\beta_1=1,$ and $\alpha_j=1,$ when $j\neq 1$. The geometric relation
can then be written as
\begin{equation}
3\beta_2\beta_3\beta_4+\beta_2+\beta_3+\beta_4=0.
\label{cond44}
\end{equation}
In the following we will also restrict our attention only to
$g(x)\propto1/x^2$, i.e.,
\begin{equation}
V(x,y)=\frac{a_1}{y^2}+\frac{a_2}{(x+\beta_2 y)^2}
+\frac{a_3}{(x+\beta_3 y)^2}+\frac{a_4}{(x+\beta_4 y)^2}.
\label{Pot4}
\end{equation}

By direct computation we obtain the leading terms of the fourth order
operator commuting with the Hamiltonian (\ref{Ham2d},\ref{Pot4}), as
\begin{align}
\IV\quad=\quad\hbar^4&[ \partial_x^4
 + 4 \partial_x \partial_y^3\beta_2\beta_3\beta_4
 -  \partial_y^4 (\beta_2 \beta_3 + \beta_2 \beta_4
 + \beta_3 \beta_4)]\nonumber\\
- 4\hbar^2&\Bigl[ \left\{\partial_x^2,g_2+g_3+g_4\right\}\nonumber\\
& -  \left\{\partial_x \partial_y, ((\beta_2+\beta_3+\beta_4) g_1
   + \beta_2 g_2 + \beta_3 g_3 + \beta_4 g_4)\right\} \nonumber\\
& -  \left\{\partial_y^2, ((\beta_2 \beta_3 + \beta_2 \beta_4
 + \beta_3 \beta_4) g_1 + \beta_2 (\beta_3+\beta_4) g_2\right. \nonumber\\
&\left.\hskip3cm + \beta_3 (\beta_2+\beta_4) g_3
 + \beta_4 (\beta_2+\beta_3) g_4)\right\}\Bigr]\nonumber\\
& + D(x,y).
\end{align}
This far we can proceed just with (\ref{cond44}).  The integrability
condition for $D$ introduces further conditions, and leads to the
following classification:

\subsubsection*{Case 1}
Assume that one $\beta$ vanishes, let us say $\beta_4=0$. Then
$\beta_3=-\beta_2$ from (\ref{cond44}) and we get two integrable
potentials: First there is the well known classical one
with $\beta_2=-\beta_3=1$
\begin{equation}
V_{4.1}=\frac{a_1}{y^2}+\frac{a_2}{(x+y)^2}+\frac{a_2}{(x-y)^2}
+\frac{a_1}{x^2}.
\end{equation}
We note that there are two degrees of freedom, all in the amplitudes.

\subsubsection*{Case 2}
There is another possibility with $\beta_3=-\beta_2$ arbitrary:
\begin{equation}
V_{4.2}=C\left(\frac1{x^2}+\frac{\beta_2^4}{y^2}\right)+
\hbar^2\left(\frac{1+\beta_2^2}{(x+\beta_2y)^2}
+\frac{1+\beta_2^2}{(x-\beta_2y)^2}-\frac1{8x^2}-\frac1{8y^2}\right).
\end{equation}
This potential has the classical limit $V=Ax^{-2}+By^{-2}$, with is
separable and therefore has a quadratic second invariant (in both
classical and quantum picture). The result above indicates that in
quantum mechanics there exists a non-separable generalization for
it. There are now two free parameters, $C$ and $\beta_2$, the solution
presented in \cite{VFC} corresponds to the choice
$C=\hbar^2\frac12(m+\frac12)^2$,
$C\beta_2^4=\hbar^2\frac12(l+\frac12)^2$

\subsubsection*{Case 3}
Next we assume that $\beta_4\neq 0$, and since due to rotational
invariance we could rotate any of the vector to the position $y$, we
may in fact assume that all $\beta$'s are nonzero (in other words that
no pair of the vectors is orthogonal). From the equations it then
follows that $a_i=\hbar^2(1+\beta_i^2)$, which normalizes the
vectors. There is one solution of this type, it is defined by
(\ref{cond44}) and
\begin{equation}
\beta_2^2+\beta_3^2+\beta_4^2-\beta_2\beta_3-\beta_3\beta_4-\beta_4\beta_2=0,
\end{equation}
which have the parameterization
\begin{equation}
\beta_2=\frac{1-a}{\sqrt{a^3-1}},\quad
\beta_3=\frac{1-a\xi}{\sqrt{a^3-1}},\quad
\beta_4=\frac{1-a\xi^2}{\sqrt{a^3-1}},
\end{equation}
where $\xi$ is a cubic root of unity $\neq 1$. This is a new result.
For real $a>1$ the potential is also real because the $g_3$ and $g_4$
terms are complex conjugates. There is only one free parameter, but
this is probably due to the special form $1/x^2$, for example if one
makes the same restriction in (\ref{V32},\ref{V32}) one can scale out $A$.

\section{Conclusions}
We have discussed the quantum integrability in two dimensions, and in
particular the existence of commuting differential operators of order
3 and 4. For the third order operator we have considered the generic
case with constant coefficients in the leading term, and found the
classical three term Calogero-Moser system (in terms of Weierstrass
functions) and two PQI (three term) systems, one of which was first
reported in \cite{VFC}. These systems also have a fourth order
commuting operator, but only for the PQI cases is the algebraic
relation between the three operators nontrivial.

It is also possible to have integrable four term potentials if the
commuting operator is of fourth order. Our analysis was here restricted
to $1/x^2$-type potential terms and we have identified two PQI
examples. Whether these potentials are also algebraically integrable is
an open question: in principle one should find for them another, still
higher order invariant, but this seems to be a formidable task.

With these results we have only scratched the surface of pure quantum
integrability. What is missing in particular is a geometric
characterization of the phenomena and comprehensive extensions to
higher orders and dimensions. The classification based on Lie algebra
root systems \cite{OP}, that works so well in the classical case, does
not include these new PQI systems.

\section*{Acknowledgments}
I would like to thank A.P. Veselov for calling my attention to this
problem, and F. van Dienen and L. Floria for pointing out further
references.


\begin{thebibliography}{9}
\bibitem{Ar} V.I. Arnold, {\it Mathematical Methods of Classical
Mechanics}, (Springer, 1978) p. 271.
\bibitem{Neu} J. von Neumann, {\it Mathematical Foundations of Quantum
Mechanics}, (Princeton UP, 1955) p. 173.
\bibitem{JH87} J. Hietarinta, {\it Direct methods for the search of
the second invariant}, Physics Reports {\bf 147}, 87-154 (1987).
\bibitem{JHq} J. Hietarinta, {\it Classical versus quantum
integrability}, J. Math. Phys. {\bf 25}, 1833 (1984).  J. Hietarinta
and B. Grammaticos, {\it On the $\hbar^2$-correction terms in quantum
mechanics}, J. Phys. A: Math. Gen. {\bf 22}, 1315-1322 (1989).
\bibitem{OO} T. Oshima, {\it Completely integrable systems with a
symmetry in coordinates}, preprint TUMS 94-6 (1994); H. Ochiai and
T. Oshima, {\it Commuting differential operators of type $B_2$},
preprint UTMS 94-65 (1994); H. Ochiai, T. Oshima and H. Sekiguchi {\it
Commuting families of symmetric differential operators}, Proc. Japan
Acad. {\bf 70 A}, 62-66 (1994); T. Oshima and H. Sekiguchi, {\it
Commuting families of differential operators invariant under the
action of a Weyl group}, J. Math. Sci. Univ. Tokyo {\bf 2}, 1-75
(1995); H. Ochiai, {\it Commuting differential operators of rank two},
preprint.
\bibitem{JHs} J. Hietarinta, {\it Solvability in quantum mechanics and
classically superfluous invariants}, J. Phys. A: Math. Gen. {\bf 22},
L143-L147 (1989).
\bibitem{BC}J.L. Burchnall and T.W. Chaundy, {\it Commutative Ordinary
Differential Operators}, Proc. Roy. Soc. (London) {\bf 118}, 557-583
(1928); ibid {\bf 134}, 471-485 (1932).
\bibitem{CV} O.A. Chalykh and A.P. Veselov, {\it Commutative Rings of
Partial Differential Operators and Lie Algebras},
Comm. Math. Phys. {\bf 126}, 597-611 (1990).
\bibitem{VFC} A.P. Veselov, M.V. Feigin and O.A. Chalykh, {\it New
integrable deformations of the Calogero-Moser quantum problem},
Russ. Math. Surveys {\bf 51}, 573-574 (1996).
\bibitem{APV} A.P. Veselov, {\it Huygen's principle and
integrability}, unpublished (1996).
\bibitem{OP} M.A. Olshanetsky and A.M. Perelomov, {\it Classically
integrable finite dimensional systems related to Lie algebras},
Phys. Rep. {\bf 71}, 313-400 (1981); M.A. Olshanetsky and
A.M. Perelomov, {\it Quantum integrable systems related to Lie
algebras}, Phys. Rep. {\bf 94}, 313-404 (1981).
\end{thebibliography}
\end{document}